\documentclass[10pt,aps,prd,twocolumn,showpacs,superscriptaddress,nofootinbib,nobibnotes,longbibliography,floatfix]{revtex4-1}

\usepackage[utf8]{inputenc}
\usepackage[T1]{fontenc}
\usepackage{comment}
\usepackage{bm}
\usepackage{mathtools,amsmath,amssymb,amsfonts,mathrsfs,eucal,graphicx,tensor,csquotes,accents,commath,chngcntr,siunitx}
\usepackage[dvipsnames]{xcolor}
\usepackage[unicode]{hyperref}
\hypersetup{colorlinks=true, citecolor=MidnightBlue,
            linkcolor=MidnightBlue, urlcolor=MidnightBlue, linktocpage=true}
\usepackage[normalem]{ulem}
\usepackage{orcidlink}

\DeclareMathAlphabet{\mathpzc}{OT1}{pzc}{m}{it}
\definecolor{darkgreen}{rgb}{0.0, 0.6, 0.0}

\newcommand{\ee}{\mathrm{e}}
\newcommand{\ii}{\mathrm{i}}
\newcommand{\dd}{\mathrm{d}}

\newcommand{\sst}{\sin^2\!\th}
\newcommand{\cct}{\cos^2\!\th}

\newcommand{\Ord}{\mathcal{O}}
\newcommand{\al}{\alpha}
\newcommand{\be}{\beta}

\newcommand{\de}{\delta}
\newcommand{\De}{\Delta}
\newcommand{\ep}{\epsilon}

\renewcommand\th{\theta}

\newcommand{\la}{\lambda}

\newcommand{\cf}{\varphi}

\newcommand{\om}{\omega}
\newcommand{\Om}{\Omega}

\newcommand{\dl}{\partial}

\newcommand{\orcid}[1]{\href{https://orcid.org/#1}{\includegraphics[width=10pt]{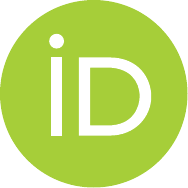}}}

\begin{document}

\author{Rajes Ghosh \orcid{0000-0002-1264-938X}}
\affiliation{
Indian Institute of Technology, Gandhinagar, Gujarat 382355, India}
\email{rajes.ghosh@iitgn.ac.in}

\author{Nicola Franchini \orcid{0000-0002-9939-733X}}
\affiliation{Université Paris Cit\'e, CNRS, Astroparticule et Cosmologie,  F-75013 Paris, France}
\affiliation{CNRS-UCB International Research Laboratory, Centre Pierre Binétruy, IRL2007, CPB-IN2P3, Berkeley, CA 94720, USA
}

\author{Sebastian H. V\"olkel \orcid{0000-0002-9432-7690}}
\affiliation{Max Planck Institute for Gravitational Physics (Albert Einstein Institute), D-14476 Potsdam, Germany}

\author{Enrico Barausse \orcid{0000-0001-6499-6263}}
\affiliation{SISSA, Via Bonomea 265, 34136 Trieste, Italy and INFN Sezione di Trieste}
\affiliation{IFPU - Institute for Fundamental Physics of the Universe, Via Beirut 2, 34014 Trieste, Italy}

\date{\today}

\title{Quasi-normal modes of non-separable perturbation equations: \\
the scalar non-Kerr case}

\begin{abstract}
Scalar, vector and tensor perturbations on the Kerr spacetime are governed by equations that can be solved by  separation of variables, but the same is not true in generic stationary and axisymmetric geometries. This complicates the calculation of black-hole  quasi-normal mode frequencies in theories that extend/modify general relativity, because one generally has to calculate the eigenvalue spectrum of a two-dimensional partial differential equation (in the radial and angular variables) instead of an ordinary differential equation (in the radial variable). In this work, we show that if the background geometry is close to the Kerr one, the problem considerably simplifies. One can indeed compute the quasi-normal mode frequencies, at least at leading order in the deviation from Kerr, by solving an ordinary differential equation subject to suitable boundary conditions. Although our method is general, in this paper we apply it to scalar perturbations on top of a
Kerr black hole with an anomalous quadrupole moment, or on top of a slowly rotating Kerr background. 

\end{abstract}

\maketitle

\section{Introduction}

Current and future measurements of gravitational waves  from mergers of binary black holes (BHs)  provide new ways to test the strong field and highly relativistic regime of general relativity (GR) \cite{LIGOScientific:2016lio,LIGOScientific:2019fpa,LIGOScientific:2020tif,LIGOScientific:2021sio}. 
One promising direction to probe the workings of gravity near BHs is to study the emission of quasi-normal modes (QNMs) in the post-merger phase of these BH binaries~\cite{1980ApJ...239..292D,Dreyer:2003bv,Berti:2005ys}. 
If GR is the correct underlying theory of gravity, then the QNM spectrum of astrophysical BHs must be uniquely determined by their mass and spin \cite{Kerr:1963ud,Carter:1971zc,Robinson:1975bv}. 
More precisely, the spectrum of a perturbed Schwarzschild BH can be obtained by solving the Regge-Wheeler \cite{Regge:1957td} (in the odd-parity sector) or Zerilli \cite{Zerilli:1970se} (in the even-parity sector) equations. 
For a perturbed Kerr BH, one has instead to solve the Teukolsky equation \cite{Teukolsky:1973ha}
(see also Ref.~\cite{PhysRevD.5.1913}).

One remarkable property of GR, which does not hold for more general gravitational theories, is that the  perturbation equations for a rotating BH can be decoupled into a radial and an angular part~\cite{PhysRevD.5.1913,Teukolsky:1973ha}. 
This separability property considerably simplifies the calculation of QNMs relative to situations in which the equations do not decouple, e.g., BH perturbations beyond GR, or even neutron star oscillations in GR. 

One obvious approach to adopt is a forward one in which QNMs are computed beyond GR on a theory-by-theory basis. 
Specific theories that have been studied (for spherical or slowly rotating BHs) include, e.g., dynamical Chern-Simons gravity~\cite{Cardoso:2009pk,Molina:2010fb,Wagle:2021tam,Srivastava:2021imr}, Einstein-dilaton-Gauss-Bonnet gravity~\cite{Blazquez-Salcedo:2016enn,Blazquez-Salcedo:2017txk,Blazquez-Salcedo:2020rhf,Blazquez-Salcedo:2020caw,Pierini:2021jxd,Pierini:2022eim}, and Lorentz-violating gravity~\cite{Franchini:2021bpt}. 
There are also recent works aiming to generalize the Teukolsky equation to more generic theories and arbitrary rotation~\cite{Li:2022pcy,Hussain:2022ins}.
This theory-by-theory approach being necessarily limited to a few case studies, much effort has also been directed at developing theory-agnostic approaches to these calculations, e.g., the parametrized QNM framework \cite{Cardoso:2019mqo,McManus:2019ulj,Kimura:2020mrh,Volkel:2022aca,Franchini:2022axs}, the effective field theory of QNMs \cite{Franciolini:2018uyq,Hui:2021cpm}, and modified perturbation equations of parametrized BH metrics \cite{Volkel:2020daa}. 
These calculations, however, are also restricted (like those of the aforementioned forward modeling approaches) to perturbations over spherically symmetric or slowly rotating BHs, and their applicability to realistic astrophysical BHs is therefore limited.
Recent theory agnostic approaches that go beyond spherical symmetry and/or the slow rotation limit are intrinsically approximate, as they adopt the eikonal limit \cite{Glampedakis:2017dvb,Glampedakis:2019dqh,Silva:2019scu,Bryant:2021xdh,Dey:2022pmv}, or simply attempt to describe possible deviations from GR in the QNM spectrum by decomposing the latter in powers of mass and spin~\cite{Maselli:2019mjd,Carullo:2021dui}. 

In this work, we outline an alternative approach that allows for efficiently computing BH QNMs for background geometries that do not yield separable perturbation equations, but which are perturbatively close to Kerr or Schwarzschild. 
Note that a similar technique has also been presented in Ref.~\cite{Cano:2020cao}, although it has been applied to rotating BHs in higher-derivative gravity. 
We present the technical details of the method in Sec.~\ref{methods}. 
To showcase its performance, we then consider the example of scalar perturbations, which we solve (in Sec.~\ref{application_results}) for a slowly rotating Kerr BH, and for a Schwarzschild/Kerr BH with an anomalous quadrupole moment. 
We stress that the latter case is highly non-trivial, as it does not yield separable perturbation equations in general. 
Our conclusions are finally presented  in Sec.~\ref{conclusions}.
Throughout this work we use units in which $G=c=1$.

\section{Methodology}\label{methods}
As discussed above, the aim of this work is to devise a general method to calculate the scalar QNMs of a stationary, axisymmetric BH spacetime given by
\begin{equation}\label{metric}
g_{\mu \nu}(r,\theta) = g_{\mu \nu}^{(0)}(r,\theta) + \epsilon\, g_{\mu \nu}^{(1)}(r,\theta)\ ,   
\end{equation}
where $g_{\mu \nu}^{(0)}$ refers to the Kerr metric, and its deviation away from GR is described by the term $g_{\mu\nu}^{(1)}$, where $\ep$ is a small perturbation parameter. The only restriction on the term $g_{\mu\nu}^{(1)}$ considered here is that it is consistent with axial symmetry. In Boyer-Lindquist coordinates $x^\mu = (t,r,\th,\cf)$ the line element of the Kerr metric is
\begin{equation}
\begin{split}\label{kerr}
    & g_{\mu \nu}^{(0)} \dd x^\mu \dd x^\nu=-\left(1-\frac{2 M r}{\rho^2}\right)\, \dd t^2 + \frac{\rho^2}{\Delta}\,  \dd r^2\\
    &+\rho^2\, \dd\th^2 +\sst 
    \left[\frac{(r^2+a^2)^2-\Delta a^2 \sst }{\rho^2}\right]\, \dd\cf^2 \\
    &-\frac{4 M r a \sst}{\rho^2}\, \dd t\, \dd\cf,
\end{split}
\end{equation}
where $\rho^2 = r^2+a^2\, \cct$, $\Delta = r^2+a^2-2 M r$, and $M$ and $a$ are the mass and the spin of the BH, respectively. The non-zero terms of $g_{\mu\nu}^{(1)}$ are unspecified functions of $r$ and $\th$.

We now study scalar perturbations $\Psi$ on top of the metric in Eq.~\eqref{metric}, satisfying the Klein-Gordon equation $g^{\mu \nu}\nabla_{\mu}\nabla_{\nu} \Psi = 0$. 
We will show that the assumption of small deviations from GR makes the perturbation equation inherit the  separability property of the Kerr case. 
We start by  expanding $\Psi$ as~\cite{Teukolsky:1973ha, PhysRevD.5.1913, Berti:2009kk}, 
\begin{equation}\label{psi}
\Psi = \int d\omega\, \sum_{\ell, m} \frac{Z_{\ell m}(r)}{\sqrt{r^2+a^2}} \, S_{\ell m} (\theta)\, \ee^{-\ii\om t+ \ii m \cf}\ .  
\end{equation}
In the case $\ep = 0$, the radial and  angular parts decouple and satisfy 
\begin{align}\label{psi0}
\frac{\dd^2 Z_{\ell m}}{\dd r_*^2} + V_{\ell m}^{(0)} (r)\, Z_{\ell m} = 0 & \, , \\
\frac{1}{\sin\th}\frac{\dd}{\dd\th} \left[\sin\th \frac{\dd S_{\ell m}(\th)}{\dd\th}\right] + & \notag \\
\bigg[ a^2\om^2\cct + \la_{\ell m} & - \frac{m^2}{\sst}\bigg] S_{\ell m}(\th) = 0 \,.\label{eq:spheroidal_scalar}
\end{align}
Here, $r_*$ is the tortoise coordinate defined by $dr/dr_* = h(r) = \De/(r^2+a^2)$, $\lambda_{\ell m}$ is a separation constant, and the effective potential is given by~\cite{Berti:2009kk},
\begin{equation} \label{V0}
    V_{\ell m}^{(0)} (r) = \frac{K^2(r)-\lambda_{\ell m}\, \Delta(r)}{(r^2+a^2)^2} -\frac{d G(r)}{dr_*} -G^2(r)\, ,
\end{equation}
where $K(r)= (r^2+a^2)\omega - a\, m$, and $G(r) = r\, \Delta(r)/(r^2+a^2)^2$. The solutions $S_{\ell m}(\th)$ of the angular equation are  the scalar spheroidal harmonics~\cite{Teukolsky:1973ha, PhysRevD.5.1913,Berti:2005gp}. They form a complete basis for angular functions and satisfy the bi-orthogonality relation with the adjoint-spheroidal harmonics $\overline{S}_{\ell m}$ defined in \cite{London:2020uva},
\begin{equation}\label{eq:completeness}
    \int \dd\Om \,  S_{\ell m}(\th)\, \overline{S}_{\ell' m'}^*(\th) \ee^{\ii (m - m') \cf} = \de_{\ell \ell'}\, \de_{m m'}\,.
\end{equation}
In the limit $a\rightarrow 0$, they reduce to scalar spherical harmonics, as $S_{\ell m}(\th)\ee^{\ii m \cf} \rightarrow Y_{\ell m}(\th,\cf)$.

In the general case $\ep\neq0$, the Klein-Gordon equation does not automatically separate, since it acquires additional terms coupling the radial and angular functions. By neglecting terms of  second order in $\ep$, the  Klein-Gordon equation becomes  
\begin{equation}\label{eomlm}
\begin{split}
\int d\omega \sum_{\ell, m} \ee^{-\ii\om t+ \ii m \cf} S_{\ell m} (\theta) \left[ \frac{d^2}{dr_*^2} + V_{\ell m}^{(0)} (r)\right] Z_{\ell m} = \epsilon\, \mathcal{J}[\Psi]\, , 
\end{split}
\end{equation}
where the source term is given by
\begin{equation}\label{J}
\mathcal{J}[\Psi] = - \frac{h(r)\rho^2}{g_{(0)}\sqrt{r^2+a^2}}\, \partial_\mu \left[ g_{(0)}\, g^{\mu \nu}_{(1)}\, \partial_\nu \Psi + g_{(1)}\, g^{\mu \nu}_{(0)}\, \partial_\nu \Psi \right]\ ,
\end{equation}
and we have used the notation $ \sqrt{-\det\, g} = g_{(0)} + \epsilon\, g_{(1)}$. Inserting the ansatz given by Eq.~\eqref{psi} in the source term, one gets the form
\begin{equation}
    \mathcal{J} = \int \dd\om \sum_{\ell, m} \ee^{-\ii\om t+ \ii m \cf} J_{\ell m }  (r,\th)
\end{equation}
where the function $J_{\ell m}$ takes the following form
\begin{equation}\label{eq:decomposition}
\begin{split}
    J_{\ell m} & = a(r, \th) Z_{\ell m}(r) S_{\ell m}'(\th) + b(r, \th) Z_{\ell m}'(r) S_{\ell m}(\th) \\
    & + c_{\ell m}(r, \th) Z_{\ell m}(r) S_{\ell m}(\th)\,.
\end{split}
\end{equation}
The explicit forms of the functions $a_{\ell m}$, $b_{\ell m}$ and $c_{\ell m}$ are given in Appendix~\ref{app:couplings}. To obtain this formula, we made use of the fact that $S(\th)$ satisfies Eq.~\eqref{eq:spheroidal_scalar} to get rid of $S''(\th)$ terms, and that, neglecting $\Ord(\ep)^2$ terms, $\dd^2 Z/\dd r_*^2$ can be eliminated with Eq.~\eqref{psi0}.\footnote{It is also worth noticing that terms proportional to $Z'(r)\, S'(\th)$ never appear due to the choice of $g_{\mu\nu}^{(1)}$ being axisymmetric.}
Now, we can make use of the fact that scalar spheroidal harmonics represent a complete basis for angular functions, meaning that it is possible to perform the following decomposition 
\begin{equation}
    J_{\ell m}(r, \th) = \sum_{\ell'} j_{\ell \ell' m}(r) S_{\ell' m}(\th)
\end{equation}
By reintroducing this form back into Eq.~\eqref{eomlm}, we have that for each term in $\ell$, $m$ it reads
\begin{equation}
    S_{\ell m}\left[ \frac{\dd^2}{\dd r^2_*} + V_{\ell m}^{(0)} \right] Z_{\ell m} = \ep \sum_{\ell'} j_{\ell \ell' m}(r) S_{\ell' m}(\th)
\end{equation}
By multiplying both sides by the adjoint spheroidal harmonic $\overline{S}^*_{\ell m}$ and using the orthonormality relation in Eq.~\eqref{eq:completeness}, one can completely decouple away the angular contribution from the radial perturbation equation. As it was singled out in Eq.~\eqref{eq:decomposition} for $J_{\ell m}$, the coefficients $j_{\ell \ell' }(r)$ must also contain terms linear in $Z$ and $Z'$, with all the possible indices $\ell'$ selected by the projection of the equation onto the spheroidal harmonics $S_{\ell m}$. Hence, without loss of generality, we have
\begin{equation}\label{eq:sourcejl}
    j_{\ell' m} \equiv j_{\ell'\ell' m} =  \alpha_{\ell' m}(r)\, Z_{\ell' m} + \beta_{\ell' m}(r)\, \frac{\dd Z_{\ell' m}}{\dd r_*}\, ,
\end{equation}
where the exact forms of $\al_{\ell' m}$ and $\be_{\ell' m}$ depend on the case of study and on the exact form of $g_{\mu\nu}^{(1)}$.
We can write the radial perturbation equation as
\begin{equation}\label{ceom}
\begin{split}
\frac{\dd^2 Z_{\ell m}}{\dd r_*^2} + \, V_{\ell m}^{(0)}(r)\, Z_{\ell m} = \epsilon\, j_{\ell m} + \epsilon\, \sum_{\ell'\neq\ell} j_{\ell' m} \, .
\end{split}
\end{equation}
One can further simplify the equation with the redefinition $Z_{\ell m} \rightarrow Z_{\ell m} \exp\left[-\ep/2\int \dd r \beta_{\ell m}(r)/h(r)\right]$ to get rid of the $ \dd Z_{\ell m}/\dd r_*$ term introduced via Eq.~\eqref{eq:sourcejl},\footnote{To ensure that the boundary conditions at $r_* \to \pm\infty$ for the new functions $Z_{\ell m}$ are the same as for the old functions, one needs to require that $\be$ go at least as $1/r_*^2$ as $r_* \to \pm\infty$.  This is verified for the examples that we consider in the following, for which $\beta(r)=0$ and therefore the transformation is not needed.} yielding
\begin{equation}\label{meom}
\frac{\dd^2 Z_{\ell m}}{\dd r_*^2} + \, V_{\ell m}(r)\, Z_{\ell m} = \epsilon\, \sum_{\ell'\neq\ell} j_{\ell' m} \, ,
\end{equation}
where the master potential is given by
\begin{equation}\label{pot}
V_{\ell m}(r) = V_{\ell m}^{(0)} (r) - \epsilon \left[ \alpha_{\ell m}(r) -\frac{1}{2}\, {\beta'}_{\ell m}(r)\, h(r) \right]\,.  
\end{equation}
Note that Eq.~\eqref{meom} still represents a coupled system
of equations between different $\ell$ modes of the radial eigenfunction $Z_{\ell m}$. Finally, with the field redefinition, $Z_{\ell m}(r) = X_{\ell m}(r)+\epsilon\, U_{\ell m}(r)$, such that $U_{\ell m}(r)$ is chosen to obey the differential equation 
\begin{equation} \label{ulm}
\frac{d^2 U_{\ell m}}{dr_*^2} + V_{\ell m}^{(0)} (r)\, U_{\ell m} = \sum_{\ell'\neq\ell} j_{\ell' m}(r) := T_{\ell m}(r_*)\,,
\end{equation}
$X_{\ell m}(r)$ must satisfy the decoupled equation
\begin{equation} \label{master}
    \frac{d^2 X_{\ell m}}{dr_*^2} + V_{\ell m}(r)\, X_{\ell m} = 0\ ,
\end{equation}
with the QNM potential $V_{\ell m}(r)$ given in Eq.~\eqref{pot}. This is an eigenvalue equation, which can be solved for the QNM frequencies $\omega_{\ell m}$ by imposing ingoing boundary conditions at the horizon and outgoing boundary conditions at infinity. 
The QNMs obtained in this way are the \textit{main result} of our perturbative calculation.

As for $U_{\ell m}(r)$, the
differential equation \eqref{ulm} has known coefficients, and the source
$T_{\ell m}$ is a functional of $Z_{\ell' m}$ (with $\ell'\neq \ell$) and their derivatives, which are known (in principle) by solving the background QNM equation given by Eq.~\eqref{psi0}. Moreover, the frequency 
$\omega=\omega_{\ell m}$ appearing in Eq.~\eqref{ulm} (through $V_{\ell m}^{(0)}$ and $T_{\ell m}$) is also known. That frequency
can be  obtained by solving either Eq.~\eqref{psi0} or Eq.~\eqref{master} as an eigenvalue problems.\footnote{The two estimates for $\omega_{\ell m}$ would differ only by ${\cal O}(\epsilon)$, and because
Eq.~\eqref{ulm}
appears already at linear order in $\epsilon$, this difference only affects the equations at ${\cal O}(\epsilon)^2$.} 
One can  solve Eq.~\eqref{ulm} by imposing that $U_{\ell m}$ must satisfy ingoing boundary conditions for $r_*\to-\infty$ and
outgoing ones for $r_*\to\infty$. 
The solution can be explicitly found by using the ``variation of parameters''-method~\cite{Nagle1993}. For this purpose, let us consider two linearly independent solutions $U_{\ell m}^{(1)}(r_*)$ and $U_{\ell m}^{(2)}(r_*)$ of the homogeneous part of Eq.~\eqref{ulm}, 
going to zero at the horizon/infinity, respectively---i.e., 
\begin{align}
    & U_{\ell m}^{(1)}(r_*)\sim \ee^{\ii \omega_{\ell m} r_*} \mbox{  for  } r_*\to -\infty\,,\label{U1}\\
    & U_{\ell m}^{(1)}(r_*)\sim A\, \ee^{\ii \omega_{\ell m}  r_*}+B\, \ee^{-\ii \omega_{\ell m}  r_*}\mbox{  for  } r_*\to +\infty\,,\\
    & U_{\ell m}^{(2)}(r_*)\sim \ee^{-\ii \omega_{\ell m}  r_*} \mbox{  for  } r_*\to \infty\,,\\
    & U_{\ell m}^{(2)}(r_*)\sim C\, \ee^{\ii \omega_{\ell m}  r_*}+D\, \ee^{-\ii \omega_{\ell m}  r_*}\mbox{  for  } r_*\to -\infty\,,\label{U2}
\end{align}
where $A$, $B$, $C$ and $D$  are constant complex coefficients.
The absence of the first derivative term $dU_{\ell m}/dr_*$ in Eq.~\eqref{ulm} implies that the Wronskian 
$(dU_{\ell m}^{(2)}/d r_*) U_{\ell m}^{(1)}-(dU_{\ell m}^{(1)}/dr_*) U_{\ell m}^{(2)}$ is a constant $W \neq 0$\footnote{Note that the Wronskian is not zero, as the two solutions are linearly independent. Should one choose instead the two solutions to satisfy  ingoing/outgoing boundary conditions at the horizon/infinity, the Wronskian would   vanish [if $\omega_{\ell m} $ is obtained
from Eq.~\eqref{psi0}] or be $\sim {\cal O}(\epsilon)$ [if $\omega_{\ell m}$ is obtained
from Eq.~\eqref{master}]. This explains why it is more convenient to choose $U^{(1,2)}_{\ell m}$ satisfying Eqs.~\eqref{U1}--\eqref{U2}.}. 
Then, we can write a solution as
\begin{align}\label{particular}
    U_{\ell m}(r_*) = & U_{\ell m}^{(1)}(r_*) \int_{r_*}^{r_{*,2}} \frac{U_{\ell m}^{(2)}(x)}{W} \, T_{\ell m}(x)\, dx \nonumber \\
    & + U_{\ell m}^{(2)}(r_*) \int_{r_{*,1}}^{r_*} \frac{U_{\ell m}^{(1)}(x)}{W} \, T_{\ell m}(x)\, dx\,,
\end{align}
with $r_{*,1}$ and $r_{*,2}$ being constants. Note that from Eq.~\eqref{eq:sourcejl}, it follows that if the functions $\alpha_{\ell'm}$ and $\beta_{\ell'm}$ remain finite as $r_*\to\pm \infty$, the source $T_{\ell m}$ diverges at most as  $\exp(\pm \ii \omega_{\ell'm} r_*)$, with $\ell'\neq \ell$, for $r_*\to\pm \infty$.\footnote{These possible divergences are simply an artefact of working in the frequency domain. Once the time dependence $\exp(-\ii \omega_{\ell'm} t)$ is restored [c.f. Eq.~\eqref{psi0}], it becomes clear that $T_{\ell m}$ is finite at future null infinity and on the future event horizon.} 
Using the behavior of $U^{(1,2)}_{\ell m}$ given in Eqs.~\eqref{U1}--\eqref{U2} and integrating by parts, it therefore follows that for $r_*\to\pm\infty$ one has $U_{\ell m}\sim \exp(\pm \ii \omega_{\ell'm} r_*) {\cal A_\pm}$, where ${\cal A_\pm}$ depends linearly on the asymptotic values of $\alpha_{\ell'm}$ and $\beta_{\ell'm}$ as $r_*\to\pm\infty$. These behaviors are sensible as they correspond to outgoing/ingoing boundary conditions for the $(\ell'\neq \ell,m)$ modes, which also appear in Eq.~\eqref{meom} due to the mode mixing term on the right-hand side.
Note, however, that the explicit form of $U_{\ell m}(r_*)$ is {\it not} needed to solve for the QNM spectrum from our master equation in Eq.~\eqref{master}, but only if one wants to reconstruct the scalar eigenfunctions. For more details we refer the reader to  Appendix \ref{App}, where we provide more algebraic details on our method.

Let us end this section with a brief summary of the methodology presented above. Given any stationary, axisymmetric BH metric $g_{\mu \nu}$, see Eq.~\eqref{metric}, perturbatively connected to the Kerr BH metric, we are interested in studying the scalar QNM modes of the spacetime. We start by expanding the scalar field $\Psi$ in scalar spheroidal harmonics in 
Eq.~\eqref{psi}, and we manipulate the Klein-Gordon equation $g^{\mu \nu} \nabla_\mu \nabla_\nu \Psi = 0$ into an inhomogeneous differential equation with a known source term, cf.~Eqs.~\eqref{eomlm} and~\eqref{J}. Then, the right hand side of Eq.~\eqref{eomlm} can itself be decomposed into spheroidal harmonics as described below Eq.~\eqref{J}. This reduces the perturbation equation to Eq.~\eqref{ceom}, with the potential $V^{(0)}_{\ell m}$ matching that of the  Kerr case [cf.~Eq.~\eqref{V0}]. Note that the right hand side of Eq.~\eqref{ceom} depends on the various radial eigenfunctions $Z_{\ell' m}$ (with $\ell' = \ell, \ell \pm 1, \ell \pm 2, \dots)$ and their derivatives. Further simplifications can be made by absorbing terms proportional to $Z_{\ell m}$ and its derivative into a redefined potential $V_{\ell m}$. 
Readers are referred to Eqs.~\eqref{meom} and~\eqref{pot} for more details.  However, the equation thus obtained still constitutes a coupled system of differential equations, due to the presence of couplings between various $\ell$ modes [via the source term in the right hand side of Eq.~\eqref{meom}]. These differential equations can be decoupled by introducing a  field redefinition, which leads to Eq.~\eqref{master}. The equation in this final form can be used to compute the QNM frequencies $\omega$.

\section{Application and Results}\label{application_results}
To illustrate the method outlined in Sec.~\ref{methods}, let us consider some examples where the $\epsilon=0$ metric $g_{\mu\nu}^{(0)}$ is that of either a Schwarzschild or a Kerr BH. For Schwarzschild, we need to set $a = 0$ and replace $e^{i m \phi}\, S_{\ell m} (\theta) \to Y_{\ell m} (\theta, \phi)$ in the expressions derived above.

\subsection{Slowly rotating Kerr BH} \label{slowrot}
Let us consider the evolution of a scalar field $\Psi$ in a slowly rotating Kerr spacetime. In this case, $g_{\mu \nu}^{ (0)}$ is a Schwarzschild metric of mass $M$ and the deviation metric $g_{\mu \nu}^{(1)}$ has only two non-zero components, $ g_{t \phi}^{(1)} = g_{\phi t}^{(1)} = - 2\, M^2\, \sin^2\theta/r$, with the expansion parameter $\epsilon = a/M$. The horizon remains at the same location as that of the background  Schwarzschild BH, i.e., at $r=2 M$. Therefore, the near horizon boundary condition also remains unchanged. Using Eq.~(\ref{J}), we obtain the source term as $\mathcal{J}[\Psi] = \frac{4 M^2 \omega\, m}{r^2 (r-2 M)}\, \Psi$, and  the final QNM equation takes the form,
\begin{equation}\label{eomSR}
\frac{d^2 Z_{\ell m}}{dr_*^2} + V_{\ell m} (r)\, Z_{\ell m} = 0\ , 
\end{equation}
with the master potential 
\begin{equation}
   V_{\ell m}(r) = V^\mathrm{Sch}_{\ell}(r) - \frac{4\, \epsilon\, \,  M^2 \omega\, m}{r^3}\,, 
\end{equation}
where 
\begin{equation}\label{vschw}
    V^\mathrm{Sch}_{\ell}(r) = \omega^2 - \frac{\ell(\ell+1)\, f(r)}{r^2} - \frac{2 M\, f(r)}{r^3}
\end{equation} and $f(r) = 1- 2 M/r$. This matches exactly the potential given in Refs.~\cite{Pani:2012bp,Pani:2013pma}. 
One then needs to solve Eq.~(\ref{eomSR}) to get the QNM frequencies for slowly rotating Kerr BHs.
 
\subsection{Schwarzschild quadrupole metric} \label{statq}
Due to the celebrated `no-hair' theorem~\cite{Carter:1971zc,Robinson:1975bv}, a vacuum BH solution of GR is characterized only by two parameters, namely the mass $M$ and spin $a$. In other words, all higher multipole moments~\cite{Hansen:1974zz, Geroch:1970cd} of the vacuum BH solutions of GR are specified uniquely by $M$ and $a$. As a consequence of this theorem, other stationary, axisymmetric and asymptotically flat vacuum GR solutions  cannot represent a regular spacetime on and outside the event horizon. See e.g.~Ref.~\cite{Manko_1992} for a specific example. However, if  there are modifications to GR, the no-hair theorem  can be violated and BHs can have extra hairs, which will in turn change the relation of various multipole moments (in particular, the quadrupole moment) with mass $M$ and spin $a$. Interestingly, such deviations from the Schwarzschild/Kerr metric have already been considered quite extensively in literature, and constraints on them from various (existing or future) astrophysical observations have been worked out (see, e.g., Refs.~\cite{Ryan:1995wh, Ryan:1997hg,Gair:2007kr, Bambi:2011jq,Bambi:2011vc, Bambi:2015ldr,Volkel:2020xlc,EventHorizonTelescope:2021dqv,Dey:2022pmv})

Here, we perform the QNM analysis of  
a Schwarzschild BH with an anomalous (i.e.~non-GR) quadrupole moment. For this purpose, let us first construct a BH metric that resembles the static (non-rotating) Hartle-Thorne metric as $r \to \infty$ \cite{1967ApJ1501005H, 1968ApJ153807H, Allahyari:2018cmg}. However, since the static Hartle-Thorne metric represents the exterior vacuum spacetime of a non-rotating star of mass $M$ and dimensionless quadrupole moment $q =\epsilon$, it cannot be extended to the location of the event horizon in a regular manner. Instead, we will construct the black hole metric to have a structure similar  to the Hartle-Thorne metric near spatial infinity (as needed to identify a quadrupole moment) and a regular Schwarzschild-like structure near the horizon.

The leading-order fall-off of the metric components for $r\to \infty$ can be obtained from the Hartle-Thorne metric given in
Eq. (2.6) of Ref.~\cite{Glampedakis:2005cf}, with the spin set to zero. We are interested in studying the effect of the quadrupole correction on the QNMs. Up to  linear order in $\epsilon$, the non-zero covariant metric components are 
\begin{align} \label{metricschq}
&g_{t t} = -f(r)\, \left[1+\epsilon\, f_1(r)\, P_{2}(\cos\theta)\right]\, , \nonumber \\
&g_{r r} = f(r)^{-1}\, \left[1-\epsilon\, f_1(r)\, P_{2}(\cos\theta)\right] \, , \\
&g_{\theta \theta} = (\sin\theta)^{-2}\, g_{\phi \phi} = r^2\, \left[1-\epsilon\, f_1(r)\, P_{2}(\cos\theta)\right]\, , \nonumber
\end{align}
where $f_1(r) = 2\, f(r)\, (M/r)^3$, and $P_{2}(\cos\theta)$ is the Legendre polynomial of second order. Our choice of $f_1(r)$ ensures that the metric components have asymptotically the right leading fall-off behaviour proportional to the quadrupole moment, namely $\epsilon/r^3 +\mathcal{O}(1/r^4)$ obtained from the functions $F_2(r,\theta)$ and $H_2(r,\theta)$ given in the Appendix of Ref.~\cite{Glampedakis:2005cf}, which is required to identify the presence of an anomalous quadrupole moment $\epsilon$. On the other hand, our construction of the metric is such that the near-horizon structure is regular and similar to the Schwarzschild case. In particular, we have chosen the functions $f_1(r)$ so that it vanishes at the horizon. Note that by construction, the event horizon location (and hence the corresponding boundary condition) remains the same as in the Schwarzschild metric.

Because of the presence of the Legendre polynomial $P_{2}(\cos\theta)$ in the metric, the source term $\mathcal{J}$ will contain the product $P_{2}(\cos\theta)\, Y_{\ell m}(\theta,\, \phi)$, i.e., a coupling between various angular momentum components and the quadrupole,
\begin{equation}
\mathcal{J}[\Psi] = \frac{4M^3\, \omega^2}{r^3}\, P_{2}(\cos\theta)\, \Psi\ .
\end{equation}
As was shown in the Methodology section, we can decompose the radial and the angular dependency in terms of $Y_{\ell' m'}(\theta,\, \phi)$, by using  recursively the relation~\cite{Pani:2013pma} 
\begin{align}\label{cY}
&\cos\theta\, Y_{\ell m} = Q_{\ell+1 m} Y_{\ell+1 m} + Q_{\ell m} Y_{\ell-1 m}, 
\end{align}
where we define $Q_{\ell m}$ and two symbols that will appear later as
\begin{gather}
   Q_{\ell m} = \sqrt{\frac{\ell^2 - m^2}{4\ell^2-1}}, \\
   A_{\ell m} = Q_{\ell m}^2 + Q_{\ell+1 m}^2, \qquad  B_{\ell m} = Q_{\ell-1 m} Q_{\ell m}.
\end{gather}
Following the method prescribed in Sec.~\ref{methods}, one finally gets the master equation,
\begin{equation} \label{Vq}
\begin{split}
\frac{d^2 Z_{\ell m}}{dr_*^2} + V_{\ell m}(r) & Z_{\ell m} = \\
3 \epsilon \, \omega^2 f_1(r)& \left( B_{\ell+2 m} Z_{\ell+2 m} + B_{\ell m} Z_{\ell-2 m}\right), 
\end{split}
\end{equation}
where $\dd r_* = \dd r /(1- 2M/r)$, the potential reads 
\begin{equation} \label{Schpot}
V_{\ell m} = V_{\ell}^\mathrm{Sch} - \epsilon\, f_1(r)\, \omega^2\, (3A_{\ell m}-1)\, ,
\end{equation}
with $V_{\ell}^\mathrm{Sch}$ being the effective potential for the Schwarzschild BH defined in Eq.~\eqref{vschw}. Then, following Sec.~\ref{methods} (for more details see Appendix \ref{App}), we can reduce the coupled system of equations to a single equation in the form of Eq.~\eqref{master}. 
\begin{figure}
\centering
\includegraphics[width=\linewidth]{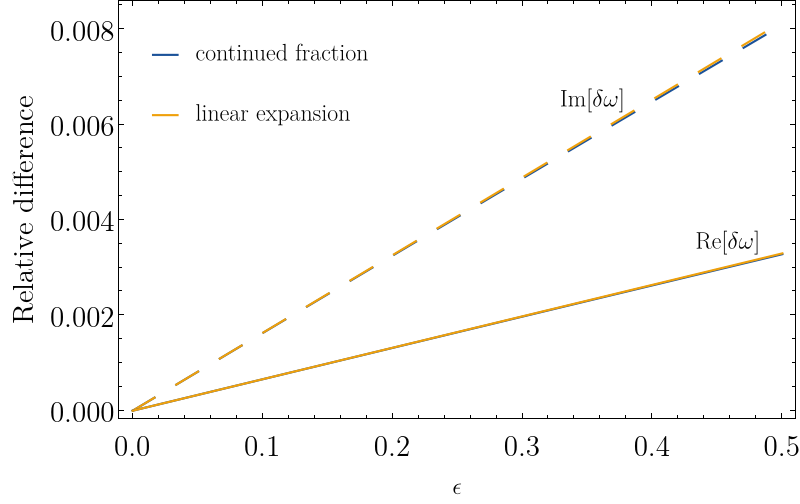}
\caption{Relative difference (absolute value) between the $\ell=m=2$ scalar mode on top of a Schwarzschild BH with quadrupolar correction $\ep$ and its GR correspondent. Solid lines refer to the real part of the mode, dashed lines to the imaginary part. We computed the modes with either a continued fraction method (blue line) or a linear expansion in $\epsilon$ (orange line).
}
\label{fig:qnm_nonrot}
\end{figure}
We have calculated the QNM frequencies from this equation using two methods, namely the method of continued fraction \cite{1985RSPSA402285L, 1986JMP271238L} and a linear expansion in $\epsilon$ along the lines of Refs.~\cite{Cardoso:2019mqo, Volkel:2022aca}. The two methods agree well, as shown in Fig.~\ref{fig:qnm_nonrot}. We have also checked the validity of the decoupling technique that we used (see Sec.~\ref{methods} and Appendix \ref{App}) by calculating QNM modes from  Eq.~\eqref{Vq}, i.e., without removing the right hand side, and these modes match well with the QNMs after decoupling Eq.~\eqref{Vq}. The coupled system was solved with the continued fraction code described in Ref.~\cite{Volkel:2022aca}. 
We computed numerically frequencies of the spectrum for a given $\ell$ and $m$, by truncating the system at some $\ell_\mathrm{max}$. 
By comparing different $\ell_\mathrm{max}$, one can study the numerical convergence of the QNMs until they achieve the desired accuracy. We found, for the cases considered, that the difference between the modes found with $\ell_\mathrm{max} = \ell + 2$ and those with $\ell_\mathrm{max} = \ell + 4$ was smaller than the error in $\ep$ given by the linear approximation. In addition, we have compared different modes of the Schwarzschild-quadrupole BH with Schwarzschild QNMs. In Fig.~\ref{fig:qnm_nonrot},  we plot the relative differences between the real and imaginary parts of these modes, for $\ell = m= 2$ and as a function of the anomalous quadrupole moment. As expected, the effect of $\epsilon$ on the QNMs is always small reflecting the fact that we are in the linear approximation regime. 

As another test, we will now discuss the ``eikonal limit'' of the QNM frequencies that we obtain. This is motivated by the fact that the eikonal limit of the QNM equation in Schwarzschild/Kerr BH spacetimes has an interesting correspondence with the circular photon orbit. In particular, one finds that in the eikonal limit the real part of the Schwarzschild/Kerr QNMs is proportional to the orbital frequency of the circular photon orbit, and the imaginary part is related to the Lyapunov exponent of the perturbed motion near the circular photon orbit \cite{Ferrari:1984zz, 1985ApJ291L33S, Iyer:1986np, Iyer:1986nq, Dolan:2010wr, Yang:2012he}. We want to check whether such correspondence is present for the case of Schwarzschild-quadrupole metric under consideration. One expects this correspondence to hold for the following reason. In the eikonal limit, we can write the perturbation as $\Psi=A\, \exp(iS)$, where $S$ is a phase factor that changes rapidly. Replacing this ansatz in the Klein-Gordon equation, we get the Hamilton-Jacobi equation $g^{\mu\nu}\partial_\mu S\, \partial_\nu S=0$ for photons under the assumption of rapidly varying phase. This implies the geodesic equation for photons~\cite{deFelice:1990hu, Barausse:2019pri}.

For this check, we focus on equatorial photon orbits. Note that because of the equatorial symmetry of the metric, initial conditions $(\theta = \pi/2,\, \dot{\theta} = 0)$ yield orbits confined in the equatorial plane. A subtle point is that the eikonal QNM equation takes the same form as the null geodesic equation when both are expressed in terms of the same tortoise coordinate. In our case, however, the QNM equation contains the Schwarzschild tortoise coordinate $r_*$, which is different from the new tortoise coordinate $\bar{r}_*$ of the Schwarzschild-quadrupole metric. On the equatorial plane, these two coordinates are related by $d\bar{r}_* = dr_*\, (1+\epsilon\, f_1(r)/2)$. By using $r_*$, the eikonal limit (obtained for $\ell=m \gg 1$) of Eq.~\eqref{Vq} is given by $d^2 X_{\ell m}/dr^2_*+ V_{\ell m}^{\textrm{eik}}(r) X_{\ell m}= 0$, where 
\begin{equation} \label{eikold}
    V_{\ell m}^{\textrm{eik}}(r) \simeq \omega^2 \big[1+\epsilon\, f_1(r)\big] -\frac{\ell^2\, f(r)}{r^2}\, .
\end{equation} 
Here, we use the fact that $Q_{\ell m} \to 0$ in the eikonal limit. When expressed in terms of new tortoise coordinate $\bar{r}_*$, the above equation generates a first derivative term $d X_{\ell m}/d\bar{r}_*$, which can then be absorbed in a redefined variable $X_{\ell m} \to \widetilde{X}_{\ell m}\, \exp\Big[-\epsilon/4\, \int d\bar{r}_*\, f(r)\, f'_1(r) \Big]$. For more details see Eq.~\eqref{tilVeik} in Appendix \ref{App2}. Thus, the final eikonal-limit QNM equation in the new $\bar{r}_*$ coordinate becomes $d^2 \widetilde{X}_{\ell m}/d\bar{r}^2_*+ \widetilde{V}_{\ell m}^{\textrm{eik}}(r)\, \widetilde{X}_{\ell m} = 0$, with the potential
\begin{equation} \label{eik}
\begin{split}
    \widetilde{V}_{\ell m}^{\textrm{eik}} \simeq  & \, \big[1-\epsilon\, f_1(r)\big] V_{\ell m}^{\textrm{eik}}(r) \simeq \\
    &\,  \omega^2 - \frac{\ell^2\, f(r)}{r^2} \big[1-\epsilon\, f_1(r)\big]\, ,
\end{split}
\end{equation}
where $V_{\ell m}^{\textrm{eik}}(r)$ is given by Eq.~\eqref{eikold}. Now, the equatorial potential for null orbits can instead be calculated as $V(r) = E^2 - L^2 f(r)\big[1-\epsilon\, f_1(r)\big]/r^2$, with $L$ and $E$ being the photon's angular momentum and energy, respectively. This potential thus matches the QNM potential in the eikonal limit (see Eq.~\eqref{eik}) provided the energy $E$ and the angular momentum $L$ of the photon are identified with the eikonal QNM frequency $\omega$ and angular momentum quantum number $\ell$.

The correspondence derived above between null geodesics and QNMs is non-trivial, but it is only valid for $\ell \gg 1$. 
One can also check whether the eikonal approximation yields
qualitatively correct results also for moderate $\ell$.
This would have implications for previous results employing the eikonal approximation in that regime (e.g. Refs.~\cite{Glampedakis:2017dvb,Dey:2022pmv,Chen:2022ynz}). To this purpose, we have numerically computed
 the QNMs in the eikonal limit  from the circular photon orbit's
 frequency $\Omega_p$ and its Lyapunov exponent $\lambda_p$ as 
\begin{align}\label{eikonal_app}
\omega_{\ell n} \approx \ell\, \Omega_p - \mathrm{i} \left(n+\frac{1}{2} \right) |\lambda_p|\,,
\end{align}
following standard procedures (see, e.g., Refs.~\cite{Glampedakis:2017dvb,Dey:2022pmv}). 
As expected, we find very good agreement with our results  for large $\ell$, but also the correct trend for moderate $\ell$. 

\subsection{Kerr- quadrupole metric} \label{kerrq}
In the same spirit of the static quadrupole metric construction of the previous section, we may design a rotating BH metric with Kerr-like structure near the event horizon at $r_+ = M + \sqrt{M^2-a^2}$, and the same leading asymptotic structure as that of the metric given in Eqs.~(2.13)--(2.16) of Ref.~\cite{Glampedakis:2005cf} at $r \gg r_+$, which is nothing but the Hartle-Thorne metric that represents the exterior vacuum spacetime of a rotating axisymmetric and stationary body of mass $M$, spin $a$, and quadrupole moment $q$~\cite{1967ApJ1501005H, 1968ApJ153807H}. The perturbative parameter in this metric is $\epsilon = q -(a/M)^2$, where $q$ is the dimensionless quadrupole moment of the BH which is not uniquely specified by mass and spin unlike Kerr. Note that by construction, the event horizon location remains same as in a Kerr spacetime.

The non-zero metric components of the metric, up to $\mathcal{O}(\epsilon)$, are therefore
\begin{equation}
\begin{split}
&g_{tt} = g_{tt}^\mathrm{Kerr} \left[1+\epsilon\, f_2(r)\, P_{2}(\cos\theta)\right] , \\
&g_{rr} = g_{rr}^\mathrm{Kerr} \left[1- \epsilon\, f_2(r)\, P_{2}(\cos\theta)\right], \\
& g_{\theta \theta} = g_{\theta \theta}^\mathrm{Kerr} \left[1- \epsilon\, f_2(r)\, P_{2}(\cos\theta)\right] , \\
& g_{\phi \phi} =  g_{\phi \phi}^\mathrm{Kerr} \left[1- \epsilon\, f_2(r)\, P_{2}(\cos\theta)\right], \\
&g_{t\phi} = g_{t\phi}^\mathrm{Kerr},
\end{split}
\end{equation}
where $f_2(r) = 2\, F(r)\, (M/r)^3$ with $F(r)= \Delta(r)/r^2$. Note that in the limit $a \to 0$, we obtain the Schwarzschild-quadrupole metric given by Eq.~\eqref{metricschq}. Let us now study the effect of the 
anomalous quadrupole moment on the QNM spectrum,  at leading (linear) order in $\epsilon$. 

To get the source term $\mathcal{J}[\Psi]$, one can directly follow the general prescription given in Sec.~\ref{methods}. However, for simplicity, we make one more approximation and expand the source term in the product $a\om$. Such expansion, while linear in $\epsilon$, is valid unless the spin is not close to the extremal Kerr limit. In this way, we can express the spheroidal harmonics $S_{\ell m}(\th)$ as a combination of generalised Legendre polynomials $P_{\ell m}(\th)$ as follows
\begin{equation}\label{SY}
    S_{\ell m} (\th) = \sum_{n=0}^{\infty} (a\om)^{2n} \sum_{k = -2n}^{N} K^{(N)}_{\ell \, k \, m} P_{\ell + k \, m}(\th)
\end{equation}
where the coefficients $K^{(N)}_{\ell \, k \, m}$ can be easily found within the Black Hole Perturbation Toolkit~\cite{BHPToolkit}.
Under this approximation, one gets the following perturbation equation,
\begin{equation} \label{VKq}
\frac{d^2 Z_{\ell m}}{dr_*^2} + V_{\ell m} (r)\, Z_{\ell m}
= \epsilon\, \sum_{\ell'\neq\ell} j_{\ell' m} +\epsilon\, \mathcal{O}(a^{N}\om^{N})\ .
\end{equation}
The term $\sum_{\ell'\neq\ell} j_{\ell' m}$ contains $Z_{\ell' m}$ and their derivatives with $\ell' = \{\ell\pm 2, \dots, \ell\pm N\}$. Then, following Sec.~\ref{methods} (for more details see Appendix \ref{App}), we can reduce the above coupled system to the decoupled equation given by Eq.~\eqref{master}, which for the case $N=2$ yields the potential, 
\begin{align} \label{Vqk}
V_{\ell m} & = V_{\ell m}^{(0)}(r)  \nonumber\\
& +3\, \epsilon\, a^2\om^4\, f_1(r)\, \left(K^{(1)}_{\ell \, 2 \, m}\, B_{\ell+2 m} + K^{(1)}_{\ell \, -2 \, m}\, B_{\ell m} \right) \nonumber \\
&- \epsilon\, f_2(r)(a^2+r^2)^{-2}\, f_3(r)
+\epsilon\, \Ord(a^3\om^3)\, ,
\end{align}
where $V_{\ell m}^{(0)}$ is given in Eq.~\eqref{V0} and  
\begin{equation} \label{KqV}
\begin{aligned}
   & f_3(r)= r\Big[\om^2\, r^3-2 m a\, \om M+ a^2\om^2 (r+2\, M)\Big](3A_{\ell m}-1)\\
    &+a^2\om^2\, \Delta(r)\Big[ (3A_{\ell m} -1)A_{\ell m}+ 3 (B_{\ell+2}^{m})^2+ 3 (B_{\ell m})^2\Big]\, .
\end{aligned}
\end{equation}
Note that $\big\{f_1,\, f_2,\, A_{\ell m},\, B_{\ell m},\, K_{\ell m}\big\}$ are defined in Secs.~\ref{statq} and \ref{kerrq}.

\begin{figure}
\includegraphics[width=\linewidth]{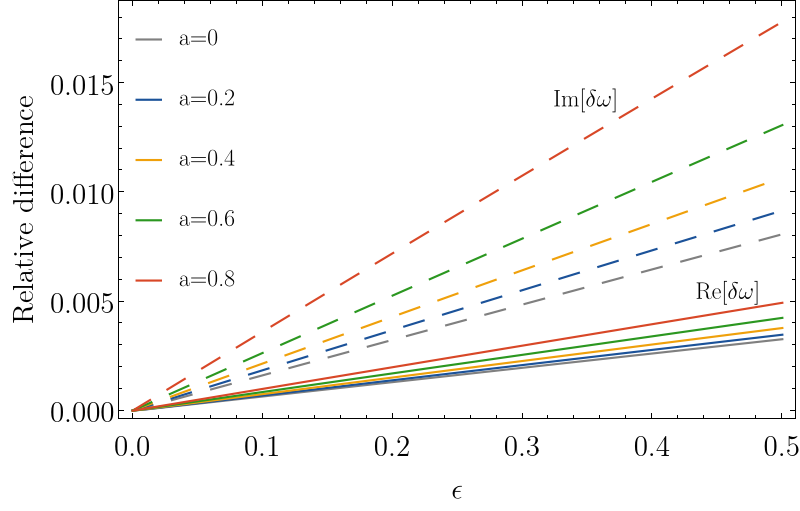}
\caption{Relative difference between the $\ell=m=2$ scalar mode on top of a Kerr BH with anomalous quadrupole moment $\ep$ and its GR correspondent, for different values of the spin $a$. Solid lines refer to the real part of the mode, dashed lines to the imaginary part. We computed the modes with the continued fraction method.}\label{fig:quadrupole_rot}
\end{figure}

\begin{figure}
\includegraphics[width=\linewidth]{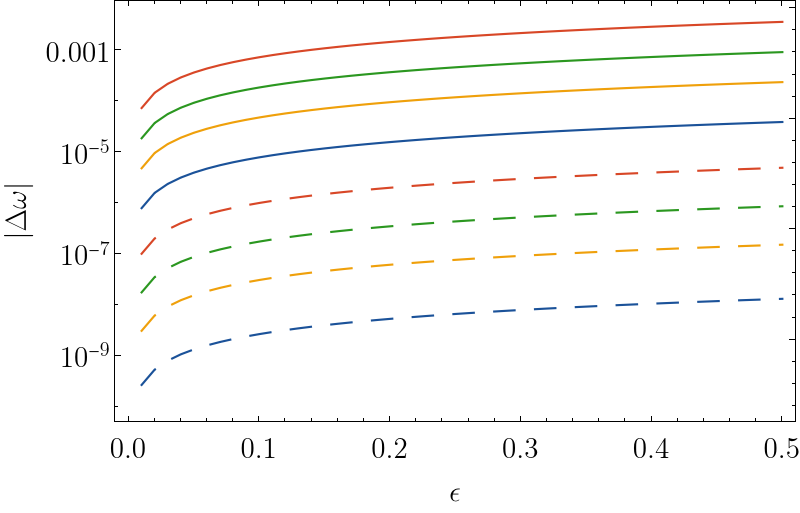} \caption{Comparison of the absolute difference for the $\ell=m=2$ scalar mode on top of a Kerr BH  with anomalous quadrupole moment $\ep$ and spin $a$. Solid lines compare the modes obtained with a $N=1$ and $N=2$ truncation in $a$ for the right-hand side of Eq.~\eqref{KqV}, whereas dashed lines show the difference between $N=2$ and $N=4$ approximations. The colors respect the same legend as Fig.~\ref{fig:quadrupole_rot}.}
\label{fig:diff_quadrupole_rot}
\end{figure}

The results of our QNM analysis are shown in Fig.~\ref{fig:quadrupole_rot} and Fig.~\ref{fig:diff_quadrupole_rot}. The relative difference between the $\ell = m = 2$ scalar QNM modes between the Kerr-quadrupole BH and a Kerr BH of same mass and spin is shown in Fig.~\ref{fig:quadrupole_rot} as a function of the anomalous quadrupole moments $\epsilon$ for different values of spins $a$. Though the relative differences in both real and imaginary parts of QNM modes increase for larger values of $a$ and $\epsilon$, they always remain small, reflecting the fact that we are working under the linear approximation. In Fig.~\ref{fig:diff_quadrupole_rot} the solid lines compare the QNM modes for the potential given in Eq.~\eqref{Vqk}, where the approximation of the source term was truncated at $N=1$, with those where terms up to $N=2$ in the $a\om$ expansion were included, whereas the dashed lines show the absolute difference between truncation at $N=4$ and $N=2$. Note that the size $|\Delta \omega|$ of the relative difference between the $\ell = m = 2$ mode obtained considering the quadratic and quartic expansions in $a\om$ of Eq.~\eqref{KqV} is smaller than the same obtained using the linear and quadratic expansions for any fixed values of $(a, \epsilon)$. It should be clear from Fig.~\ref{fig:diff_quadrupole_rot} that the expansion in the spin presented here is made under different form than in the pure slow-rotation expansion~\cite{Kojima:1992ie,Pani:2012bp,Pani:2012vp}.

In general, it is non-trivial to demonstrate the explicit correspondence between  QNMs and the unstable photon orbit in the eikonal limit for arbitrary spacetimes, as we did in the previous section for the Schwarzschild quadrupole case. 
The correspondence for the Kerr-Newman BH has been demonstrated recently \cite{Li:2021zct}. 
In order to test our analytic framework for the Kerr quadrupole case, we numerically computed the QNMs in the 
eikonal limit via Eq.~\eqref{eikonal_app}, as we did for the Schwarzschild quadrupole metric, and compared them to our results.
Again,  we found very good agreement for large $\ell$, as expected, but also the correct trend for moderate $\ell$. 

\section{Conclusions}\label{conclusions}

In this work, we outlined a procedure that allows one to compute BH QNMs in cases where the separability of the perturbation equations is not achievable exactly, but the background spacetime's deviation from the Kerr solutions is small. 
The underlying idea, which has also been used in the case of rotating BHs in higher-derivative gravity in Ref.~\cite{Cano:2020cao}, is to rewrite the perturbation equations in an approximate fashion by making use of the spheroidal harmonics basis. 
While the resulting system of equations shows couplings between radial functions with different quantum numbers,  as a result of the deviations from Kerr, we find that the system can be diagonalized (i.e. decoupled) via a suitable redefinition of the radial functions.
As implicit in our perturbative treatment, we find that  the corrections to the QNM spectrum of Kerr/Schwarzschild BHs are small, which allows for mapping our results into the parameterized QNM framework 
of Refs.~\cite{Cardoso:2019mqo,McManus:2019ulj,Volkel:2022aca}. 

The consistency of our method has been checked in several ways. 
We verified that when applied to a slowly rotating Kerr BH, one obtains the known analytic result for the (scalar) perturbation equation. 
We also quantified the accuracy of the QNM frequencies when computed at different orders when expanding the spheroidal harmonics into spherical harmonics, and we found good agreement. 
Furthermore, we computed the QNM frequencies with a continued fraction method, and we verified that the results agree with the linear order of the parametrized QNM framework of Refs.~\cite{Cardoso:2019mqo, McManus:2019ulj, Volkel:2022aca}. 
Finally, we also used the eikonal approximation relating the orbital frequency at the photon ring and its Lyapunov exponent to the QNMs. We have found very good agreement for large $\ell$, and the correct trend also for moderate $\ell$. 

In this paper, we have focused on scalar perturbations on top of non-Kerr/non-Schwarzschild metrics as a non-trivial toy problem.
The extension to the full gravitational case requires one to choose a specific theory of gravity, compute the background metric, and then derive the set of perturbation equations. We will tackle this extension in future work. While our paper was under review, a preprint where the method is applied to the study of gravitational perturbations in higher-derivative gravity appeared~\cite{Cano:2023tmv}.

\acknowledgments
We would like to thank P.~Pani and A.~Maselli for insightful comments on a draft of this work and for many enlightening conversations on quasi-normal modes. 
We would also like to thank Lionel T.~London for an important remark on the completeness of the spheroidal harmonics. 
The research of R.~G.~is supported by the Prime Minister Research Fellowship (PMRF ID: 1700531), Government of India. R.~G.~also wants to thank all members of SISSA and IFPU for their warm hospitality and pleasant
company over his stay there, during which the most part of this work was completed. 
S.V. acknowledges funding from the Deutsche Forschungsgemeinschaft (DFG) - project number: 386119226. 
E.B. acknowledges support from the European Union's H2020 ERC Consolidator Grant ``GRavity from Astrophysical to Microscopic Scales'' (Grant No.  GRAMS-815673). This work was also supported by the EU Horizon 2020 Research and Innovation Programme under the Marie Sklodowska-Curie Grant Agreement No. 101007855. This work makes use of the Black Hole Perturbation Toolkit.

\appendix

\section{The source term}\label{app:couplings}
In this appendix, we provide the explicit forms of the coefficients $a$, $b$ and $c_{\ell m}$ appearing in Eq.~\eqref{eq:decomposition}. The modification to the metric $g_{\mu\nu}^{(1)}$ is chosen to be the most general stationary and axisymmetric configuration (not necessarily circular)~\cite{Wald:1984rg}.
We have
\begin{align}
    & a(r,\th) = \dl_\th \left[ v_2 - u \frac{a^2 r^2 \sst}{\rho^2 \De} \right] + 2 f_{t\th}, \\
    & b(r,\th) = \De \,\dl_r \left[ v_1 - u \frac{a^2 r^2 \sst}{\rho^2 \De} \right] + 2 f_{tr}, \\
    & c_{\ell,m}(r,\th) = r \, \om \, w_{10} \left[ \frac{K}{\De} + r \om  \right] + a^2 \om^2 w_{02} \cct \notag \\
    & \quad -\frac{4M a r u F}{\rho^2 \De} \left[ m \left(\rho^2 -r\right) + a r \om \sst \right]  \notag \\
    & \quad + \frac{m^2 w_{23}}{\sst} + w_{12} \la_{\ell m} \notag  + \dl_r f_{tr} \\
    & \quad + a m w_{13} \frac{2M r \om - a m}{\De} + \frac{1}{\sin\th}\dl_{\th}\left(\sin\th f_{t\th} \right),
\end{align}
where we define
\begin{align}
    u(r,\th) & = \frac{1}{2}\left( f_{tt} - 2 f_{t \cf} + f_{\cf\cf} \right), \\
    v_0(r,\th) & = \frac{1}{2} \left( - f_{tt} + f_{rr} + f_{\th\th} + f_{\cf\cf} \right), \\
    v_1(r,\th) & = \frac{1}{2} \left( f_{tt} - f_{rr} + f_{\th\th} + f_{\cf\cf} \right), \\
    v_2(r,\th) & = \frac{1}{2} \left( f_{tt} + f_{rr} - f_{\th\th} + f_{\cf\cf} \right), \\
    v_3(r,\th) & = \frac{1}{2} \left( f_{tt} + f_{rr} + f_{\th\th} - f_{\cf\cf} \right) ,
\end{align}
and $w_{ij} = v_i - v_j$. The functions are defined as
\begin{align}
    & f_{tt}(r,\th) =g_{tt}^{(1)} / g_{tt}^{(0)}, \qquad f_{rr}(r,\th) = g_{rr}^{(1)} / g_{rr}^{(0)} , \\
    & f_{\th\th}(r,\th) = g_{\th\th}^{(1)} / g_{\th\th}^{(0)} , \qquad f_{\cf\cf}(r,\th) = g_{\cf\cf}^{(1)} / g_{\cf\cf}^{(0)} , \\
    & f_{t\cf}(r,\th) = g_{t\cf}^{(1)} / g_{t\cf}^{(0)}, \qquad f_{tr}(r,\th) = \De F g^{(1)}_{tr}, \\
    & f_{t\th}(r,\th) = F g^{(1)}_{t\th} , \qquad F = \ii \left( \om + \frac{r K}{\rho^2 \De} \right).
\end{align}

\section{Decoupling the wave equation} \label{App}
In Sec.~\ref{methods}, we have discussed a general method to decouple a system of coupled QNM equations of the form given in Eq.~\eqref{meom}. Thus, one may directly use this method to deal with the QNM equations as given by Eqs.~\eqref{Vq} and~\eqref{VKq}. 

However, in this appendix, we will use a slight variation of the method presented in Sec.~\ref{methods}, in order to show that our results reduce to those of Ref.~\cite{Pani:2013pma} for some limiting cases. For this purpose, consider a system of coupled wave equations of the  form
\begin{align} \label{Zeq}
\frac{d^2 Z_{\ell m}}{dr_*^2} + V_{\ell m} (r)\, Z_{\ell m} =\, &\epsilon\, \Big[ f_{\ell m}^{(1)}(r)\, Z_{\ell+2, m} \nonumber \\
& + f_{\ell m}^{(2)}(r)\, Z_{\ell-2, m}\Big]\ .
\end{align}
Note that the right hand side is a special case of the more general situation shown in Eq.~\eqref{meom}. Now, let us assume that the ratio $\Big( f_{\ell m}^{(1)}/f_{\ell m}^{(2)} \Big)$ is $r$-independent, which is the case for Eq.~\eqref{Vq}. This motivates us to introduce a field redefinition, $X_{\ell m}(r_*) = Z_{\ell m}(r_*) + \epsilon\, \tilde{Z}_{\ell m}(r_*)/n(r)+ \epsilon\, U_{\ell m}(r_*)$, where $r_*$ is the background tortoise coordinate defined by $dr/dr_* = h(r)$ and $ \tilde{Z}_{\ell m} = c_{\ell m}\, Z_{\ell+2,m} - d_{\ell m}\, Z_{\ell-2,m}$. We want to choose the $r$-independent coefficients $(c_{\ell m}, d_{\ell m})$ and the function $U_{\ell m}$ in such a way that $X_{\ell m}$ satisfies the standard wave equation in decoupled form when $\mathcal{O}(\epsilon^2)$ quantities are neglected, see Eq.~\eqref{master}. One such choice is given by
\begin{align} \label{cdu}
&c_{\ell m} = \frac{f_{\ell m}^{(1)}(r)\, n(r)}{V_{\ell+2,m}^{(0)} - V_{\ell m}^{(0)}},\quad d_{\ell m} =  \frac{f_{\ell m}^{(2)}(r)\, n(r)}{V_{\ell m}^{(0)}-V_{\ell-2,m}^{(0)}}; \\
\nonumber \\ 
&\frac{d^2 U_{\ell m}}{dr_*^2} + V_{\ell m}^{(0)} (r)\, U_{\ell m} = \partial_{r_*} \Big( \frac{n'(r)\, h(r)}{n^2(r)}\Big)\, \tilde{Z}_{\ell m} \nonumber \\
& \kern10em + \frac{2n'(r)\, h(r)}{n^2(r)}\, \frac{d \tilde{Z}_{\ell m}}{dr_*}\ .
\end{align} 
where $\tilde{Z}_{\ell m}$ on the right hand side of the above equation must be considered as $\tilde{Z}_{\ell m}(\epsilon = 0)$ and $V_{\ell m}^{(0)}$ is the Schwarzschild (Kerr) QNM potential given in Eq.~\eqref{vschw} (Eq.~\eqref{V0}). To check  our result, one can readily see that the choice $n(r) =1$ and $U_{lm} = 0$ agrees with the case presented in Ref.~\cite{Pani:2013pma} for massless scalar perturbations of a slowly rotating Kerr BH.

A few comments are in order here. First, we should choose $n(r)$ in such a way that $(c_{\ell m},d_{\ell m})$ are $r$-independent. Given that the ratio $\Big( f_{\ell m}^{(1)}/f_{\ell m}^{(2)} \Big)$ is $r$-independent, that is only possible if the ratio $\Big(V_{\ell+2, m}^{(0)} - V_{\ell m}^{(0)}\Big)/\Big(V_{\ell m}^{(0)}-V_{\ell-2,m}^{(0)}\Big)$ is also $r$-independent. In fact, this is the case for the Schwarzschild metric 
with an anomalous quadrupole moment 
which we  considered in this paper, see Eq.~\eqref{Vq}. In that case, it is easy to see that $n(r) = r$, $d_{\ell m} = -3M^3 \omega^2 B_{\ell}^{m}/(2\ell-1)$, and $c_{\ell m} = d_{\ell+2,m}$ represent a valid solution. In contrast, if at least one of the ratios $\Big( f_{\ell m}^{(1)}/f_{\ell m}^{(2)} \Big)$ and $\big(V_{\ell+2,m}^{(0)} - V_{\ell m}^{(0)}\big)/\big(V_{\ell m}^{(0)}-V_{\ell-2,m}^{(0)}\big)$ depends on $r$, the coefficients of $Z_{\ell+2,m}$ and $Z_{\ell-2,m}$ in the field redefinition cannot both be set proportional to $n^{-1}(r)$. 

After finding $\big(n(r), c_{\ell m}, d_{\ell m}\big)$, we can substitute their values in the differential equation for $U_{\ell m}$ given in Eq.~\eqref{cdu}, where both the source and the frequencies appearing in the potential are in principle known.
Therefore, we can  always solve for $U_{\ell m}$, cf.~Eq.~\eqref{particular} and related discussion. As a result, the final equation takes the form given in Eq.~\eqref{master}, which can be solved to find  the QNM frequencies $\omega$.

One can use the above method repeatedly for decoupling an equation with higher order mode couplings too, where the $\ell$ mode is coupled to $\{\ell-\bar{\ell}, \dots ,\ell-2, \ell-1, \ell+1, \ell+2, \dots, \ell+\bar{\ell}\}$ modes, see for example Eq.~\eqref{VKq} in the Kerr-quadrupole case discussed in Sec.~\ref{kerrq}. By symmetry, it is easy to see that higher multipole perturbations (such as quadrupole, octupole and so on) of the background BH metric may give rise to such couplings. Therefore, from a phenomenological point of view, one may directly start with a potential $V_{\ell m} = V_{\ell m}^{(0)} + \sum_{i} \epsilon_i V_{\ell m}^{i(1)}$ in Eq.~(\ref{master}), where the background $V_{\ell m}^{(0)}$ has been perturbed by contributions $V_{\ell m}^{i(1)}$ coming from various higher multipoles. This motivates  the work presented in Ref.~\cite{Volkel:2022aca}. It is not hard to concoct a similar method for decoupling an equation with a source term containing derivatives of $Z_{\ell' m}$ as well. \\

\section{Wave equation in terms of $\bar{r}_*$} \label{App2}
In this appendix, we want to address one more issue that may arise. In Eq.~\eqref{master}, the derivatives are with respect to the Kerr/Schwarzschild tortoise coordinate $r_*$ which may differ from the tortoise coordinate $\bar{r}_*$ of the metric given by Eq.~(\ref{metric}). However, if the horizon location  remains the same, we may still choose to work with the old tortoise coordinate, as the near-horizon (ingoing) boundary condition remains unaltered~\cite{Berti:2009kk, Pani:2013pma}. However, for a general scenario where this is not the case, we have to tackle the problem of incorporating the QNM boundary conditions properly.

For this purpose, let us assume that the new and old tortoise coordinates are related by $d\bar{r}_* = dr_*\, [1+\epsilon\, g(r)]$, where $g$ depends  on the radial coordinate only. Note that one can always put the function $g$ in such a form by writing the metric in  coordinates in which the radial coordinate is constant on the horizon. Using this relation, we can express the derivatives in Eq.~(\ref{master}) with respect to $\bar{r}_*$, to get
\begin{align} 
\frac{d^2 X_{\ell m}}{d\bar{r}_*^2} + \, \epsilon\, h(r)\, g'(r)\, \frac{d X_{\ell m}}{d\bar{r}_*}+ V_{\ell m}(r) \Big(1- & 2\, \epsilon\, g(r)\Big) X_{\ell m} \nonumber \\
+ \mathcal{O}(\epsilon^2)= 0\ ,
\end{align}
where $dr/dr_* = h(r)$. We can omit the first derivative term on the left-hand side by a field redefinition, $X_{\ell m} \to \widetilde{X}_{\ell m}\, \exp\Big[-\epsilon/2\, \int d\bar{r}_*\, h(r)\, g'(r) \Big]$. Then, the final master equation becomes
$d^2 \widetilde{X}_{\ell m}/d\bar{r}_*^2 + \widetilde{V}_{\ell m}(r)\, \widetilde{X}_{\ell m} = 0$, with the redefined potential as 
\begin{equation} \label{tilW}
 \widetilde{V}_{\ell m}(r) = V_{\ell m}(r)\Big[1-2\, \epsilon\, g(r)\Big] - \frac{1}{2}\, \epsilon\, h(r)\, \frac{d \big[h(r)\, g'(r) \big]}{dr}\, .
\end{equation}
As a demonstration of the method above, let us discuss a suggestive example, where we consider the effect of a small  charge on the Schwarzschild scalar QNMs. In other words, the spacetime is Reissner-Nordstr\"{o}m (RN)
with a small charge $|Q| \ll M$, and the location of the horizon is different from Schwarzschild. We will consider $\epsilon = Q^2$ as our perturbative parameter. In this case, $g(r) = -\big[ r^2\, f(r) \big]^{-1}$ and $h(r) = f(r)$. Then, Eq.(\ref{pot}) gives the potential\footnote{Since for the case of a RN BH, the function $g(r)$ and the $\epsilon$-dependent term in $V_{\ell}(r)$ given by Eq.~\eqref{RNrs} diverge at $r=2M$, our perturbative analysis is not strictly valid near $r=2M$. However, following our analysis, one gets the correct (and regular) result at the end: see Eq.~\eqref{RN}.}
\begin{equation} \label{RNrs}
V_{\ell}(r) = V_{\ell}^\mathrm{Sch}(r) - \epsilon\, \frac{2\, M^2 + 2\, r^4\, \omega^2 -r^2\, f(r)\, (\ell^2+\ell-1)}{r^6\, f(r)}\ .
\end{equation}
Therefore, using Eq.~\eqref{tilW} the master QNM equation becomes
\begin{equation}
\frac{d^2 \widetilde{X}_{\ell m}}{d\bar{r}_*^2} + \widetilde{V}_{\ell}(r)\, \widetilde{X}_{\ell m} = 0\ ,
\end{equation}
where the master potential is
\begin{equation} \label{RN}
    \widetilde{V}_{\ell}(r) = V_{\ell}^\mathrm{Sch}(r) - \epsilon\, \frac{6\, M + (\ell+2)(\ell-1)\, r }{r^5}\, , 
\end{equation}
which can be verified by a direct calculation of QNM equation for a RN spacetime.

Another interesting conclusion follows directly from Eq.~\eqref{tilW}. Note that in the eikonal limit ($\ell = m \gg 1$), the last term (which is $\ell$-independent) on the right hand side of Eq.~\eqref{tilW} can be neglected and the eikonal potential takes a very simplified form:
\begin{equation} \label{tilVeik}
    \widetilde{V}_{\ell m}^{\textrm{eik}}(r) \simeq V_{\ell m}^{\textrm{eik}}(r)\Big[1-2\, \epsilon\, g(r)\Big]\, ,
\end{equation}
where $V_{\ell m}^{\textrm{eik}}(r)$ is the eikonal limit of the potential $V_{\ell m}(r)$ given in Eq.~\eqref{pot}. We will use this equation in Sec.~\ref{statq} (see Eq.~\eqref{eik}) to derive the eikonal QNM potential in the Schwarzschild-quadrupole case.

\bibliography{literature}

\end{document}